\documentclass[%
 reprint,
 amsmath,amssymb,
 aps,
]{revtex4-1}
\usepackage{graphicx}% Include figure files
\usepackage{dcolumn}% Align table columns on decimal point
\usepackage{bm}% bold math
\usepackage[outdir=./]{epstopdf}
\usepackage{comment}
\usepackage[caption=false]{subfig}
\newcommand{\mb}{\mathbf}
\begin{document}

%\preprint{APS/123-QED}

\title{Dynamical Transitions in a Dragged Growing Polymer Chain}% Force line breaks with \\
%\thanks{}%

\author{A. Malek}
 %\altaffiliation{Institut f. theor. Physik, University of G\"ottingen}%Lines break automatically or can be forced with \\
\author{R. Kree}%
 \email{kree@theorie.physik.uni-goettingen.de}
\affiliation{Institut f. Theoretische Physik, Universit\"at G\"ottingen, Friedrich-Hund Pl. 1, 37077 G\"ottingen, Germany\\}


%\affiliation{
%\affiliation{
%\affiliation{%


\date{2016/05/15}% It is always \today, today,

\begin{abstract}
We extend the Rouse model of polymer dynamics to situations of non-stationary chain growth. For a dragged polymer chain of length $N(t)=\gamma t^\alpha$, we find two transitions in conformational dynamics. At $\alpha=1/2$, the propagation of tension and the average shape of the chain change qualitatively, while at $\alpha=1$ the average center-of-mass motion stops. These transitions are due to a simple physical mechanism: a race duel between tension propagation and polymer growth. Therefore they should also appear for growing semi-flexible or stiff polymers. The generalized Rouse model inherits much of the versatility of the original Rouse model: it can be efficiently simulated and it is amenable to analytical treatment. 
\end{abstract}

\pacs{Valid PACS appear here}% PACS, the Physics and Astronomy
\maketitle

%\tableofcontents

%
%
%
The dynamics of polymer conformations in space and the dynamics of polymer reactions are both well established topics, which have been extensively studied in physics and chemistry, albeit as independent of each other in most cases. Theoretical reaction models are usually formulated as stochastic processes in the space of chemical sequences  of monomers without refering to the chain conformation \cite{odian}, whereas standard models of conformational dynamics are set up for non-reacting chains of fixed length \cite{doi-edwards}. However, if reactions occur over time scales on which the chain conformation changes considerably, the two aspects of polymer dynamics may become coupled in several nontrivial ways. Cyclization reactions \cite{cyclization} and polymer ratchets \cite{polymer-ratchet}  provide examples of such couplings. Polymer transport through nanopores has been studied as the dynamics of dragged shrinking polymers \cite{sakaue1, sakaue2}. 
We studied another simple yet generic type of conformation-reaction coupling: the conformational dynamics of a chain during its growth under non-stationary conditions.  
This situation is frequently met in biological as well as technically important systems.
In cellular environments, monomers or regulators of reactions are provided in limited amount and degrade or diffuse away from the reaction volume and may lead to decreasing reaction rates \cite{calcium, signalling, gactindiffuse, gactinbiomimetic, myo1c}. Polymers like F-actin undergo polymerization on time scales of significant conformational changes \cite{Factin,actintimecourse1, actintimecourse2, factintimeresolve1}.
The same scenario also applies to living polymerization \cite{living-polymer} in technical applications \cite{living-poly-applications}. Despite its widespread appearance, the consequences of non-stationary polymerization on conformations have not been investigated in detail.  
To address this problem, we have extended the well-known Rouse model of conformational dynamics \cite{rouse} to situations of ongoing chain growth. 
In this letter we introduce our model and use it to show that a dragged polymer growing with decreasing velocities exhibits two observable dynamical transitions by varying the rate of decrease.
   
The basic setting is depicted in Fig.1a.  For the purpose of illustration, we consider the simple case of irreversible chain-growth polymerization at one end of a flexible chain, while the other end is dragged by a constant force $\mb{F}$.   During its growth the  polymer's length $\hat{N}(t)$ increases by $1$ at random reaction times, whereas the average chain length $N(t)$ changes according to the rate equation
\begin{equation}
	\frac{dN}{dt}=\lambda(t)\rho(t)	
	\end{equation}   
with reaction rate $\lambda$ and monomer concentration $\rho$.	
For monotonically decreasing $\lambda\rho=o(t^{-1})$, the polymer length will approach a finite value  $N(t\to \infty)$, and the conformational dynamics will approach that of a chain of this fixed length. In the following we  are interested in situations with $\lambda\rho=O(t^{\alpha-1})$  ($\alpha > 0$), so that $N=O(t^\alpha)$ is growing without limit with a growth velocity, which is either decreasing  ($\alpha<1$), constant  ($\alpha=1$) or increasing ($\alpha>1$). To be specific we restrict the discussion to pure power law growth  $N(t)=\gamma t^{\alpha}$. 

Let us now briefly introduce our model. 
We start from the Rouse dynamics because its linear structure allows us to find exact results for averaged quantities directly at the transitions and to give a simple, reliable approximate solution for all values of $\alpha$. Nevertheless we will argue that the dynamical transitions found in the simple model will also appear for stiff or semi-flexible chains. 
Rouse dynamics \cite{rouse} is based on a standard equilibrium statistical description of flexible polymer chains, which consists of $N+1$ point-like quasi-monomers at positions $\mb{R}_i $ $(i=0,\cdots N)$, coupled by harmonic spring forces $\mb{f}_i=k\mb{b}_i$, proportional to the bond vectors $\mb{b}_i=\mb{R}_{i}-\mb{R}_{i-1}$, and with coupling constant $k=3k_BT/b^2$ of entropic origin at temperature $T$ and Kuhn length $b$. 
It is formulated
as a set of linear Langevin equations  $\zeta\dot{\mb{R}}_i=-k(2\mb{R}_{i}-\mb{R}_{i+1}-\mb{R}_{i-1})+\mb{F}\delta_{i,0}+\pmb{\eta}_i(t)$ for the $\mb{R}_i$, with friction coefficient $\zeta$ and white noise of thermal origin, characterized by vanishing average and correlations $\langle\eta_\alpha(t)\eta_\beta(t')\rangle=2\zeta k_BT\delta_{\alpha\beta}\delta(t-t')$ between cartesian components $\eta_\alpha,\; \alpha=x,y,z$.
These Langevin equations hold for all monomers $i=0,\cdots N$, if we introduce two phantom monomers $\mb{R}_{-1}=\mb{R}_0$ and $\mb{R}_{N+1}=\mb{R}_{N}$. 
For our purposes, we have also introduced a constant dragging force acting on the first monomer, $\mb{F}_i=\mb{F}\delta_{i,0}$.  In the following,  we use the Kuhn length $b$, the monomer diffusion time
$\tau_0=k/\zeta$ and the coupling $k$ to de-dimensionalize all our quantities and equations without changing denotations of variables.  
We model chain-growth polymerization reactions as inhomogeneous continuous time random walks \cite{ctrw-montroll}. For a chain, which is growing at one end the process is characterized by the joint PDF of waiting times and added bond vectors, $\Psi(\mb{b},t,\tau)=p(\mb{b})\psi(t,\tau)$. The added bond vector $\mb{b}=\mb{b}_{\hat{N}+1}=R_{\hat{N}(t)+1}-R_{\hat{N}(t)}$ is drawn from an isotropic Gaussian distribution with vanishing average and equilibrium bond length $\langle \mb{b}^2\rangle= b^2$. 
$\psi(t, \tau)d\tau$ is the probability that after a reaction at time $t$, the next reaction occurs at $t+\tau$ within $d\tau$. It is determined by the propensity $\lambda(t)\rho(t)$ \cite{gillespie}. We compare our simulation results to  analytical calculations within a continuum version of the Rouse model, which is based on the partial differential equation (PDE)
$\partial_t\mb{R}(s,t)=\partial_s^2\mb{R}(s,t)+\pmb{\eta}(s,t)$
for the position vector field $\mb{R}(s,t)$.    
Here $0\leq s\leq \hat{N}(t)$ replaces the discrete chain index $i$ and 
$(2\mb{R}_{i}-\mb{R}_{i+1}-\mb{R}_{i-1})$ becomes the second partial derivative $\partial_s^2\mb{R}(s,t)$. Chain growth and drag force enter the continuum model via boundary conditions
$\partial\mb{R(0,t)}/\partial s=\mb{b}(0,t)=\mb{f}$ and $\partial\mb{R}(\hat{N}(t),t)/\partial s= \mb{b}(\hat{N}(t),t)=\pmb{\xi}(t)$, where $\pmb{\xi}(t)$ is the vector-valued point process of added bonds, characterizing the noisy growth. 

\begin{figure}[htp]

\subfloat[]{%
  \includegraphics[clip,width=0.8\columnwidth]{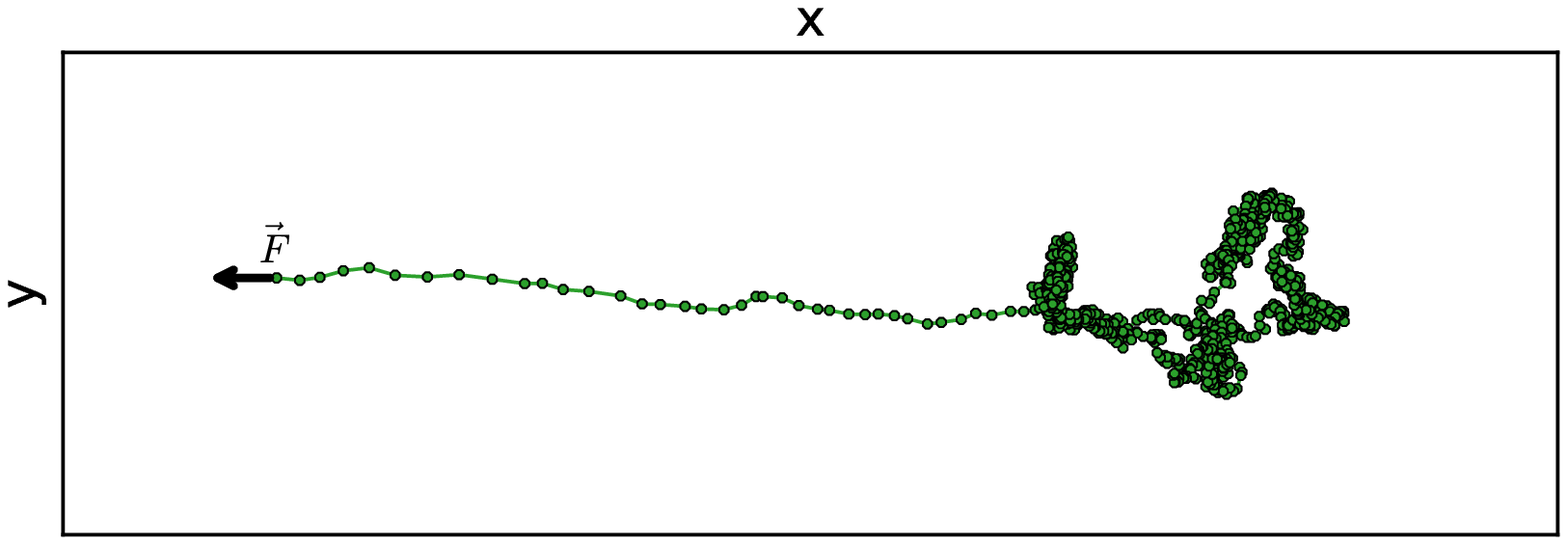}%
}

\subfloat[]{%
  \includegraphics[clip,width=0.8\columnwidth]{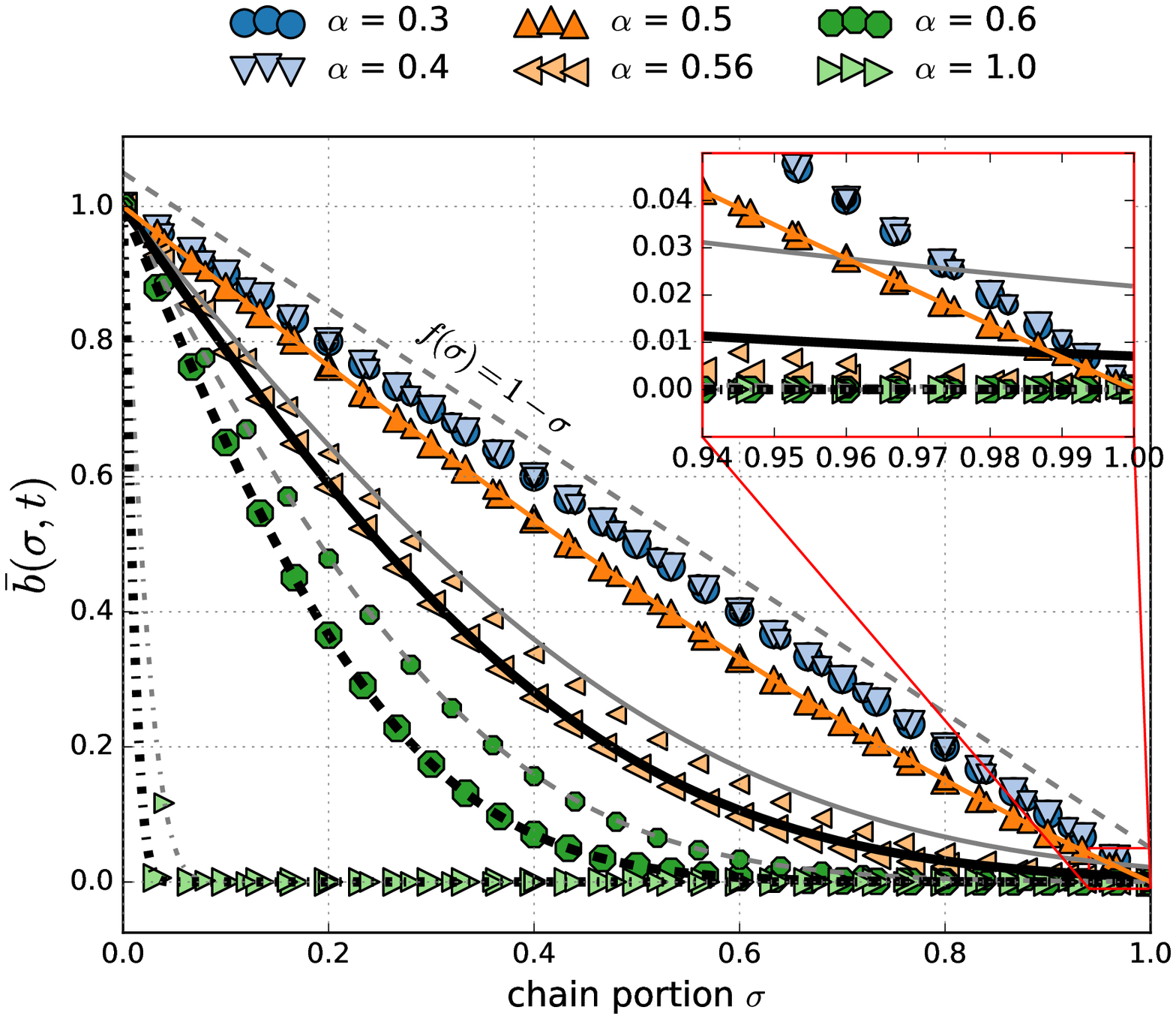}%
}

\caption{(a)Snapshot at $N=10^4$ of a growing chain dragged by a unit force $\mb{F}$ in $-x$  direction with growth exponent $\alpha=0.7$. Whereas conformations of non-growing chains would be linearly stretched (slightly displaced light dashed line marked $1-\sigma$), growing chain conformations  with $\alpha>1/2$ possess stem-flower like shapes, with  randomly coiled and streched parts. (b) distribution of average tension $\bar{b}_x$  along the chain at $N=10^3$ (small markers) and $N=10^4$ (big markers) for different values of $\alpha$ vs. normalized monomer position $\sigma=s/N$ in the chain. The lines have been obtained from analytical calculations and approximations described in the main text. The inset of a magnified region near $\sigma=1$ depicts the error of the similarity solution Eq.(\ref{eq:similarity}) for $N=10^3$ (light line) and $N=10^4$ (bold line) at $\alpha=0.65$. Also shown is the analytical solution at $\alpha=0.5$  \label{fig:TENSION_VS_SIGMA}}.
\end{figure}

We now turn to a discussion of the two dynamical transitions, which we found by increasing $\alpha$: one in  long-time propagation of tension and in average shape of the conformation  at $\alpha=1/2$  and  one in the  center-of-mass motion at $\alpha=1$. 
Fig.\ref{fig:TENSION_VS_SIGMA}(b)  shows the averaged tension $|\overline{b}_x(\sigma, t)|$  vs. $\sigma=s/N$ for $N=1000$ and $N=10000$ at different values of $\alpha$. The chain is dragged by a unit force pointing in -x direction. For $\alpha<1/2$ the tension reaches the linear limit distribution $1-\sigma$ as expected for a chain of fixed length. 
For $\alpha>1/2$ the tension  piles up near the dragged end $\sigma=0$, while an increasing fraction of the chain forms a nearly force-free coil. Exactly at $\alpha=1/2$, the tension relaxes towards a non-linear limit distribution $\overline{b}_x(\sigma)$ along the chain. 
To understand the physical origin of this transition, suppose that the dragging starts at $t=0$, so that the initial tension is localized at $s=0$. It spreads diffusively along the chain and within a time $t>0$, it will build up tensions in bonds with $s\sim O(t^{1/2})$. If $\alpha<1/2$, tension propagates faster than polymer growth and will reach the growing end even for longest times. However, if $\alpha>1/2$ polymer growth is faster than the propagation of tension and the fraction of stretched bonds decreases with time. Thus the transition is the result of a racing duel between the tension and the polymer growth. For stiff polymers tension propagation has been calculated in \cite{tensionSemiFlexI} to proceed sub-diffusively as $t^{1/4}$ for longest times, and as $t^{1/8}$ for shorter times, so we expect transitions at corresponding values of $\alpha$.   

\begin{figure}[h]
\centering
\includegraphics[scale=0.4]{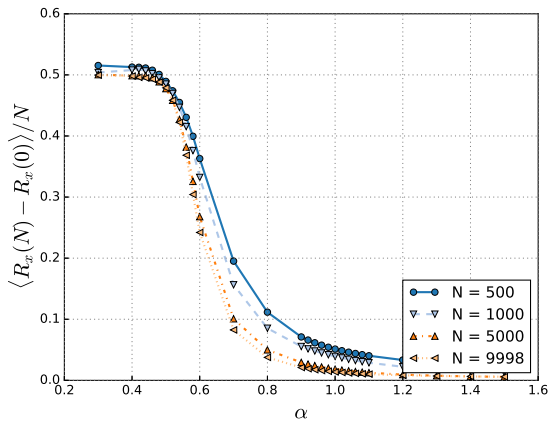}
\caption{End-to-end distance along the direction of external force, divided by chain length $N$. For $\alpha<0.5$ it scales like $N$. Beyond $\alpha=0.5$, it decreases sharply, because an increasing fraction of the chain becomes an unstretched random coil. \label{fig:R_TIP_DIF_R_0_VS_ALPHA}}
\end{figure}

\begin{figure}[h]
\centering
\includegraphics[scale=0.4]{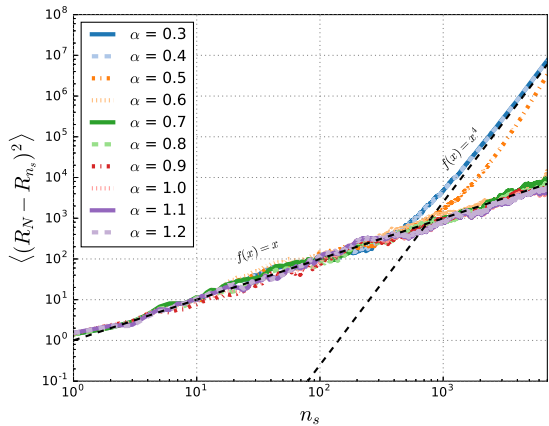}
\caption{Squared distance of monomer at chain position $s=N-n_s$ from the tip position $\mb{R}(N(t),t)$ vs $n_s$ for $N=10^4$.  For $\alpha>1/2$ the influence of the dragging force is restricted to a very thin boundary layer. For $\alpha<1/2$ the coiled part of the chain shrinks to a boundary layer (of a few hundred monomers) near the tip . For $\alpha=1/2$, the stretched part of the chain does not follow a simple power law scaling. \label{fig:R_EE_2_R}}
\end{figure}

Figs. (\ref{fig:R_TIP_DIF_R_0_VS_ALPHA}) and (\ref{fig:R_EE_2_R}) quantify the coiled and stretched parts of the chain and indicate the development towards a sharp transition with increasing time and chain length. Fig. (\ref{fig:R_TIP_DIF_R_0_VS_ALPHA}) shows the average length of the stem flower shape $L_x=\langle (R_x(N(t))-R_x(0))\rangle$ in the direction of force, divided by chain length $N(t)$ vs. $\alpha$. For $\alpha<1/2$ this length is a finite fraction of $N$, whereas for $\alpha>1/2$ $L_x/N$ should drop to zero. The width of the transition region decreases with increasing time leaving a step discontinuity for $t\to \infty$.   
Fig. (\ref{fig:R_EE_2_R}) shows the average of the squared distance between the position $\mb{R}(N(t),t)$ of the growing tip and the position $\mb{R}(s,t)$ of the s'th monomer vs. $N(t)-s$ for $N(t)=10000$ and for different $\alpha$. For all values $\alpha<1/2$, the crossover from stretched to coiled conformations is obvious, whereas for all $\alpha>1/2$ nearly the whole chain remains unstretched. The qualitative differences in temporal development of the statistics of internal distances below and above $\alpha=1/2$ offers ways to observe the transition by FRET and PET techniques \cite{fret, fret-pet}.  


\begin{figure}[h]
\centering
\includegraphics[scale=0.4]{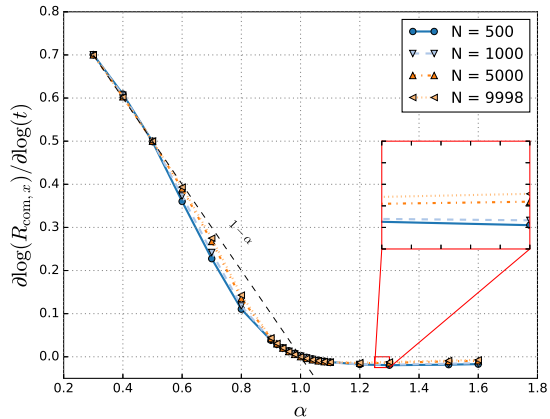}
\caption{Logarithmic derivative of the logarithm of x-component of center-of-mass vs. $\alpha$. The dashed line corresponds to the simple expectation $t^{1-\alpha}$ as explained in the main text. The inset shows the approach to zero (upper boundary of inset) for $1.25<\alpha<1.3$ \label{fig:COM}}.
\end{figure}

If $\alpha$ is increased to and beyond $\alpha=1$, the average motion of the center of mass position $\overline{\mb{R}}_{cm}(t)$ stops, as shown in Fig.(\ref{fig:COM}).    
In a chain of fixed length it is obvious that $\overline{\mb{R}}_{cm}(t)=\overline{\mb{R}}_{cm}(0)-\mb{f}t/N$. In the growing chain, the motion of $\overline{\mb{R}}_{cm}$ is coupled to the motion of the tip position $\overline{\mb{R}}(N(t),t)$ by the equation 
\begin{equation}
\frac{d}{dt}[N(t)\overline{\mb{R}}_{cm}]=\frac{dN}{dt}\overline{\mb{R}}(N(t),t)-\mb{f}.
\end{equation}
For $\alpha>1/2$ we expect from the discussion above that the tip performs an essentially unbiased random walk. We checked that $\overline{\mb{R}}_x(N(t),t)$ approaches a constant and thus expect $\overline{\mb{R}}_{cm}\to \mb{C} -\mb{f}t/N(t)$ for $\alpha>1/2$. Fig.(\ref{fig:COM}) shows that for $1/2<\alpha<1$ there are deviations from this simple expectation  up to $N=10^4$, for which we  cannot offer a simple physical explanation at present.  

Finally, we discuss our numerical methods and the analytical results we achieved.
For chains of fixed length an important reason for the widespread use of the Rouse model is its simple and exact solvability. In our simulations, we also made use of this feature.   
The coupled random system of growth and thermal motion can in principle be simulated by numerically solving the Langevin equations of conformational dynamics in between two reaction events, which are drawn according to $\Psi(\mb{b},t,\tau)$. But if the chain growth slows down (for $0<\alpha<1$), a straightforward Brownian dynamics simulation becomes inefficient at long times. We then use the analytically calculated transition PDFs $p(\{\mb{b}\}, t+\tau\; |\; \{\mb{b}'\},t)$ between conformations $\{\mb{b}\}=\{\mb{b}_0,\cdots \mb{b}_{\hat{N}(t)}\}$. This allows for an exact simulation of Rouse dynamics with freely chosen time steps between two growth events. 
Our simulations started with a dumbbell ($\hat{N}_0 = 1$), with bond vector $\mb{b}_1 = (1,0,0)$. We  averaged over $100$ to $1000$ samples, until statistical errors were reduced to allow for smooth interpolations between data points.

The analytical continuum model is a random, inhomogeneous moving boundary value problem (MBVP), which makes it hard to treat without further approximations. 
To make progress we neglected chemical noise and thus replaced the random chain length $\hat{N}(t)$ by $\overline{N}(t)$. The quality of this approximation is checked by direct comparison with simulation results. For the purpose of this work, we only need the average $\overline{b}$ of bond vectors over thermal and chemical noise, which possess the deterministic Dirichlet boundary conditions  
  $\overline{\mb{b}}(N(t),t)=\mb{0}$ and  $\overline{\mb{b}}(0,t)=\mb{f}$. 
 The deterministic MBVP  is mapped to a fixed BVP by introducing $\sigma =s/N(t)$ as independent variable. This leads to 
 \begin{equation}
 \partial_t \overline{\mb{b}}(\sigma,t) =\frac{\dot{N}}{N} \sigma \partial_\sigma \overline{\mb{b}}(\sigma,t) + \frac{\partial^2_\sigma \overline{\mb{b}}}{N(t)^2},
 \label{eq:landautransf}
 \end{equation}
 which has to be supplemented by $\overline{\mb{b}}(0,t)=\mb{f},\; \overline{\mb{b}}(1,t)=\mb{0}$.  
The remaining dimensionless parameter $\gamma$ in $N(t)=\gamma t^\alpha$ can be eliminated by rescaling $t$ and $s$  in the form $\tilde{t}=\gamma^{2 \mu}t$ and $\tilde{s}=\gamma^{\mu}s$ with $\mu=1/(2\alpha-1)$, so that the rescaled variable $\tilde{s}$ runs from $0$ to $\tilde{t}^\alpha$. The force $\mb{f}\delta(s)$ then changes accordingly to $\gamma^{-2\mu}\mb{f}\delta(\gamma^{-\mu}\tilde{s})$, i.e. $\tilde{\mb{f}}=\gamma^{-\mu}\mb{f}$. After rescaling notation is  changed back to original denotations. Within this scaled model, the force is the only remaining model parameter.  
The prefactors of the drift term and the diffusion term in Eq.(\ref{eq:landautransf}) decay in time like $\dot{N}/N\sim t^{-1}$ and $N^{-2}\sim t^{-2\alpha}$. For $\alpha<1/2$ the diffusive term is dominant for long times. After introducing a new time variable $\tau$ by $N^2 d/dt=d/d\tau$ the drift term becomes a regular perturbation of the diffusion equation, which may be treated by standard methods. The leading term of the perturbation expansion is just the Rouse solution in rescaled variables $\sigma$ and $\tau$, and therefore it approaches $\overline{\mb{b}}\to\mb{f}(1-\sigma)$ for long times.
 
At the transition points $\alpha=1/2$ and $1$, the model without chemical noise can be solved exactly. 
At $\alpha=1/2$ both $\dot{N}/N$ and $N^{-2}$ become $1/t$. Multiplying  Eq.(\ref{eq:landautransf}) by $t$ and introducing $\tau=\ln t$ as new time scale, the problem is mapped to the stationary form $\partial_\tau \overline{\mb{b}}=\mb{H}_{OU}\overline{\mb{b}}$, with $\mb{H}_{OU}=\sigma\partial_\sigma+\partial^2\sigma$ denoting the well-known Fokker-Planck operator of an Ornstein-Uhlenbeck process on a finite interval. The eigenfunctions can be expressed by parabolic cylinder functions, but the stationary solution is obtained by elementary integration. For large $t$, we find that 
\begin{equation}
	\overline{\mb{b}}(\sigma, t)\to \mb{f}\left[1 - \text{erf}(\sigma/2)/\text{erf(1/2)}\right],
	\end{equation}   
This limit is shown in Fig(\ref{fig:TENSION_VS_SIGMA}). 

At $\alpha=1$ conformational properties can be calculated from the exact Green's function of Eq.(\ref{eq:landautransf}), which we find to be of the form
\begin{equation}
\mathcal{G}(\sigma,\tau|\sigma',\tau')=e^{\Phi(\sigma,\tau)}G(\sigma,\tau|\sigma',\tau')e^{-\Phi(\sigma', \tau')},
\end{equation}
where $G(\sigma,\tau|\sigma',\tau')$ is the Green's function of the free diffusion equation and 
$\exp[\Phi(\sigma,t)]=N^{-1/2}(t)\exp[-N(t)\sigma^2/4]$. Further details of this exactly solvable case will be given elsewhere.

Except for $\alpha=1/2$ and $\alpha=1$ we do not have the general solution of the BVP Eq.(\ref{eq:landautransf}). However, any localized initial condition will spread diffusively, and its influence on the growing end of the chain will therefore decrease with time for $\alpha>1/2$. 
Thus we may approximate the growing Rouse chain by a half infinite chain to calculate the propagation of disturbances.  For example,  we find that the similarity solution 
\begin{equation}
\bar{\mb{b}}(\sigma,t) = \mb{f}\left[1- \text{erf}\left( \sigma\frac{N(t)}{2\sqrt{t}}\right)\right]
\label{eq:similarity}
\end{equation}
of Eq.(\ref{eq:landautransf}) provides a good global approximation for  $\alpha>1/2$ and for nearly all values of $\sigma$ outside of a shrinking region close to $\sigma=1$. 
The solution matches the boundary condition at $\sigma =0$  exactly, and it rapidly approaches the correct boundary value at $\sigma = 1$.  The inset of Fig(\ref{fig:TENSION_VS_SIGMA}) illustrates the difference between the approximate solution and our simulation results.  

In conclusion, we have introduced a model for the dynamics of conformations of a dragged growing Rouse chain with $N(t)\propto t^\alpha$. We found a transition in the propagation of tension around $\alpha=1/2$ and an arrest of center-of-mass motion at and beyond $\alpha=1$. As the transitions are based on a simple race duel between tension and polymer growth, we expect corresponding transitions in semi-flexible or stiff polymers. The analytical solutions at $\alpha=1/2$ and $\alpha=1$ and the approximate similarity solution Eq.(\ref{eq:similarity}) allow for a more detailed analysis of growing chains, which will be discussed further in a separate publication. The model can easily be applied to other scenarios, like, for example, a growing chain fixed in space at one or both ends, a chain in external linear flow or relaxation of sudden stresses in growing chains.  It can also be applied to shrinking chains, which provides an alternative starting point to the problem of translocation of a polymer chain through a nanopore \cite{sakaue1, sakaue2}.  Thus it extends the versatility of the original Rouse model to the non-equilibrium regime of living polymers.  

We gratefully acknowledge support from SFB 937 \textit{Collective behavior of soft and biological matter}.

\bibliography{PRL_BIB}% Produces the bibliography via BibTeX.
\end{document}